\documentclass[prb,reprint,twocolumn,floatfix,twoside,tbtags,showpacs]{revtex4-1}

\usepackage[utf8]{inputenc}
\usepackage{graphicx}
\usepackage{subfigure}
\usepackage{amsmath,amssymb}
\usepackage{bm}	
\usepackage{xspace} 

\newcommand{\etal}{\textit{et al}.\@\xspace}
\newcommand{\ie}{\textit{i.e.}\@\xspace}
\newcommand{\cf}{\textit{cf}.\@\xspace}

\newcommand{\md}{\mathrm{d}}

\begin{document}

\title{Dislocation core field. II. Screw dislocation in iron}

\author{Emmanuel \surname{Clouet}}
\email{emmanuel.clouet@cea.fr}
\affiliation{CEA, DEN, Service de Recherches de Métallurgie Physique,
91191 Gif-sur-Yvette, France}

\author{Lisa \surname{Ventelon}}
\affiliation{CEA, DEN, Service de Recherches de Métallurgie Physique,
91191 Gif-sur-Yvette, France}

\author{F. \surname{Willaime}}
\affiliation{CEA, DEN, Service de Recherches de Métallurgie Physique,
91191 Gif-sur-Yvette, France}


\date{\today}
\begin{abstract}
	The dislocation core field, which comes in addition to the Volterra elastic field, is studied for the $\langle111\rangle$ screw dislocation in alpha-iron. This core field, evidenced and characterized using \emph{ab initio} calculations, corresponds to a biaxial dilatation, which we modeled within the anisotropic linear elasticity. We show that this core field needs to be considered when extracting quantitative information from atomistic simulations, such as dislocation core energies. Finally, we look at how dislocation properties are modified by this core field, by studying the interaction between two dislocations composing a dipole, as well as the interaction of a screw dislocation with a carbon atom.

\end{abstract}
\maketitle

\section{Introduction}

\emph{Ab initio} calculations have revealed that a $\langle111\rangle$
screw dislocation in $\alpha$-iron creates a core field in addition to 
the Volterra elastic field \cite{CLO09a}.
This core field corresponds to a pure dilatation in the $\left\{111\right\}$
plane perpendicular to the dislocation line. 
It is responsible for a non negligible volume change per unit length of dislocation line.
The core field decays more rapidly than the Volterra field, 
as the displacement created by this core field varies as
the inverse of the distance to the dislocation line, whereas the displacement
caused by the Volterra field varies as the logarithm of this distance.
Such a dislocation core field is not specific to iron: 
a similar dilatation induced by the core of the screw dislocation can be deduced
from the analysis of core structures obtained
from first-principles in other body-centered cubic (bcc) metals, such as Mo and Ta \cite{ISM00,FRE03}. 
Atomistic simulations in other crystal structures have also
led to such a core field \cite{GEH72,HOA76,WOO77,HEN04,HEN05}.

The purpose of this paper is to characterize this core field
for a $\langle111\rangle$ screw dislocation in iron, and to see 
how it modifies the dislocation properties.
In that purpose, \emph{ab initio} calculations have been used to 
obtain the dislocation core structure. Then the dislocation core field 
has been modeled within the linear anisotropic elasticity theory,
using the approach initially developed by Hirth and Lothe \cite{HIR73},
and generalized in the preceding paper \cite{CLO11a} to incorporate
the core field contribution into the dislocation elastic energy.
This modeling allows extracting quantitative information 
from \emph{ab initio} calculations, such as the dislocation core energy. 
Finally, the effect of the core field on the interaction 
of a screw dislocation
with a carbon atom has been investigated,
as well as on the properties of a screw dislocation dipole.

\section{Atomistic simulations}

\subsection{Dislocation dipole}

\begin{figure}[btp]
	\begin{center}
		\includegraphics[width=0.99\linewidth]{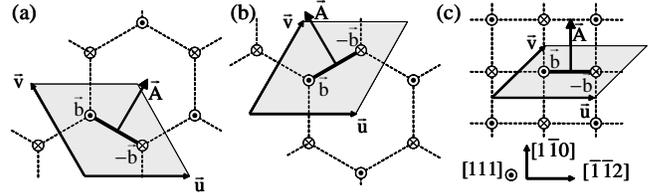}
	\end{center}
	\caption{Screw dislocation periodic arrangements used for \emph{ab initio}
	calculations: (a) T and (b) AT triangular arrangements, (c) quadrupolar 
	arrangement. 
	$\mathbf{u}$ and $\mathbf{v}$ are the unit cell periodicity vectors, 
	and $\mathbf{A}$ the dipole cut vector.
	In all cases, the Burgers vector $\mathbf{b}=\frac{1}{2}a_0\left[ 111 \right]$ for easy,
	and $-\frac{1}{2}a_0\left[ 111 \right]$ for hard core configuration.}
	\label{fig:PBC}
\end{figure}

Fully periodic boundary conditions have been selected to study 
the $\langle 111 \rangle$ screw dislocation in $\alpha$-Fe. 
A dislocation dipole is introduced into the simulation box,
using three different periodic distributions of dislocations.
Within the triangular arrangement, initially proposed by Frederiksen and Jacobsen \cite{FRE03},
the dislocations are positioned on a honeycomb network [Figs.~\ref{fig:PBC}(a)
and \ref{fig:PBC}(b)] that strictly preserves the threefold symmetry of the bcc lattice 
along the $\left[111\right]$ direction.
Two different variants, which are linked by a $\pi/3$ rotation,
are possible for this periodic arrangement: 
the twinning (T) [Fig.~\ref{fig:PBC}(a)] and the anti-twinning (AT)
[Fig.~\ref{fig:PBC}(b)] triangular arrangement.
The name of the variant refers to the fact that the dislocation dipole
has been created by shearing a $\left\{112\right\}$ plane either
in the T or AT orientation.
The third periodic arrangement, represented in Fig.~\ref{fig:PBC}(c),
is equivalent to a rectangular array of quadrupoles.
The periodicity vectors $\mathbf{u}$ and $\mathbf{v}$ defining these different arrangements
are given in Ref.~\onlinecite{VEN07}.
For all periodic arrangements, the periodicity vector along the dislocation
line is taken as the minimal allowed vector, \ie the Burgers vector 
$b=\frac{1}{2}a_0\left[ 111 \right]$ where $a_0$ is the lattice parameter.
Thus the number of atoms in the simulation 
box is directly proportional to the surface $S$ of the 
unit cell perpendicular to the dislocation line.

For each periodic arrangement, the dislocations are positioned 
at the center of gravity of three neighboring $\left[ 111 \right]$ 
atomic columns. 
Depending on the sign of the Burgers vector, two types of cores
can be obtained \cite{VIT74}.
In the easy core configuration, the helicity of the lattice is locally reversed compared 
to the helicity of the perfect lattice. 
This configuration has been found to be the most stable in all atomistic simulations
in bcc transition metals.
In the hard core configuration, the three central neighboring $\left[ 111 \right]$ 
atomic columns are shifted locally so that these atoms lie in the same $\left( 111 \right)$
plane.
This configuration actually corresponds to a local maximum 
of the energy of the dislocation arrangement.
It can be nevertheless stabilized numerically by symmetry in the atomistic simulations.
The three periodic arrangements
sketched in Fig.~\ref{fig:PBC}
allow us to simulate 
a dipole whose both dislocations are either in their easy or hard 
core configuration\cite{VEN07}.

The dislocation dipole is introduced into the simulation cell 
by applying the displacement field of each dislocation, 
as given by the anisotropic linear elasticity.
A homogeneous strain is also applied to the periodicity
vectors, to minimize the elastic energy contained within the simulation box
\cite{LEH98,CAI01,CAI03,LI04,CLO09a}.
This homogeneous strain corresponds to the plastic strain 
produced when the dislocation dipole is introduced into the simulation 
unit cell.
It is given by
\begin{equation*}
	\varepsilon_{ij}^0 = -\frac{b_i A_j + b_j A_i}{2S}.
\end{equation*}
The orientations of the Burgers vector, $\mathbf{b}$, 
and of the dipole cut vector, $\mathbf{A}$, are defined in Fig.~\ref{fig:PBC}.
Then the atomic positions are relaxed so as to minimize the energy of the simulation
box computed by \emph{ab initio} calculations \cite{VEN07}. 

Two types of simulations have been performed in this work: 
simulations at constant volume and at zero stress. 
One can keep the periodicity vectors fixed and minimize the energy 
only with respect to the atomic positions. 
Within this constant volume simulation, the simulation box is subject to a homogeneous stress. 
Within the zero stress simulation, the unit cell is allowed to relax its size and shape,
so that the homogeneous stress vanishes at the end of the relaxation.
In both cases, the dislocation core field can be identified.

\subsection{\emph{Ab initio} calculations}

The present \emph{ab initio} calculations in bcc
iron have been performed in the density functional theory (DFT) framework 
using the \textsc{Siesta} code\cite{SOL02}, \ie the pseudopotential approximation and 
localized basis sets, as in Refs.~\onlinecite{FU04}
and~\onlinecite{VEN07}. 
Comparison with plane-waves DFT calculations \cite{VEN10} has shown 
that this \textsc{Siesta} approach is reliable to study dislocations in bcc iron.
The charge density is represented on a real space grid with 
a grid spacing of $0.06$\,{\AA}
that has been reduced after self-consistency to $0.03$\,\AA.  
The Hermite-Gaussian smearing technique with a $0.3$\,eV width 
has been used for electronic density of state broadening. 
These calculations are spin-polarized and eight valence electrons 
are considered for iron. 
The Perdew-Burke-Ernzerhof (PBE) generalized gradient approximation (GGA) scheme is used 
for exchange and correlation. 
A $3\times3\times16$ $k$-point grid is used for the dislocation calculations
with unit cells containing up to 361 atoms,
and a $16\times16\times16$ grid for the elastic constants.

The obtained Fe lattice parameter is $a_0=2.88$\,\AA, 
in good agreement with the experimental value (2.85\,\AA).
The DFT elastic constants are deduced from a fit on a fourth 
order polynomial over the energies for different strains
ranging from $-2$ to $2$\,\%.
This leads to the values of $248$, $146$ and $69$\,GPa
for the elastic constants $C_{11}$, $C_{12}$ and $C_{44}$ respectively,
expressed in Voigt notation in the cubic axes.
These values are close to the experimental ones, $C_{11}=243$ 
and $C_{12}=145$\,GPa, except for the shear modulus $C_{44}$
which is found stiffer experimentally (116\,GPa). 
As a consequence, the elastic anisotropy within DFT
is less pronounced than experimentally: DFT calculations lead
to an anisotropic ratio $A=2C_{44}/(C_{11}-C_{12})=1.35$ instead of 2.36.

The three \emph{ab initio} elastic constants 
yield to a shear modulus in $\{110\}$ planes
$\mu_{110} = \left( C_{11}-C_{12}+C_{44} \right)/3$
equal to 57\,GPa, which is close to the experimental value, 
$\mu_{110}=71$\,GPa. 
This parameter is of major importance for $\left<111\right>$ dislocations
as it controls their glide in $\{110\}$ planes.
Another important quantity is the logarithmic prefactor $K=\mu b^2 / 4\pi$
controlling the main contribution to dislocation elastic energy.
For a screw $\left<111\right>$ dislocation, the shear modulus appearing
in this prefactor is equal to 56\,GPa for \emph{ab initio} data 
and 64\,GPa for experimental ones, thus in close agreement. 
The error between \emph{ab initio} and experimental elastic constants
should therefore not affect too much our results.
For consistency, all elastic calculations below are performed 
using \emph{ab initio} elastic constants.

\section{Core field characterization}

The dislocation core field can be modeled by an equilibrium 
distribution of line forces\footnote{One can also consider 
a distribution of dislocation dipoles to model the core field. 
We found that the core field representation as a line force distribution
works better for the $\protect\langle111\protect\rangle$ screw dislocation in iron,
and therefore we did not consider any dislocation dipole in the core field.}
parallel to the dislocation and located close to its core \cite{GEH72,HIR73}. 
For the $\langle111\rangle$ screw dislocation in iron, the center
of this distribution corresponds exactly to the position of the dislocation, 
\ie to the center of gravity of three $\langle111\rangle$ neighboring atomic 
columns, for symmetry reasons. 
At long range, and at a point defined by
its cylindrical coordinates, $r$ and $\theta$, this distribution generates an elastic displacement 
given by a Laurent series (see preceding paper \cite{CLO11a})
\begin{equation*}
	\mathbf{u}(r,\theta) = \sum_{n=1}^{\infty}{\mathbf{u}_n\frac{1}{r^n}}.
\end{equation*}
The main contribution of this series, \ie the term $n=1$, is completely 
controlled by the first moments $M_{ij}$ of the line force distribution. 
Knowing this second rank tensor $M_{ij}$, one can not only predict
the  elastic displacement and stress associated with the core field \cite{HIR73},
but also the contribution of the core field to the elastic energy
and to the dislocation interaction with an external stress field
\cite{CLO11a}. 
It is thus important to know the value of the first moment tensor $M_{ij}$,
and we will see how it can be deduced from \emph{ab initio} calculations.

\subsection{Simulations with fixed periodicity vectors}
\label{sec:fixed_periodicity_vectors}

\begin{figure}[btp]
	\begin{center}
			\includegraphics[width=0.95\linewidth]{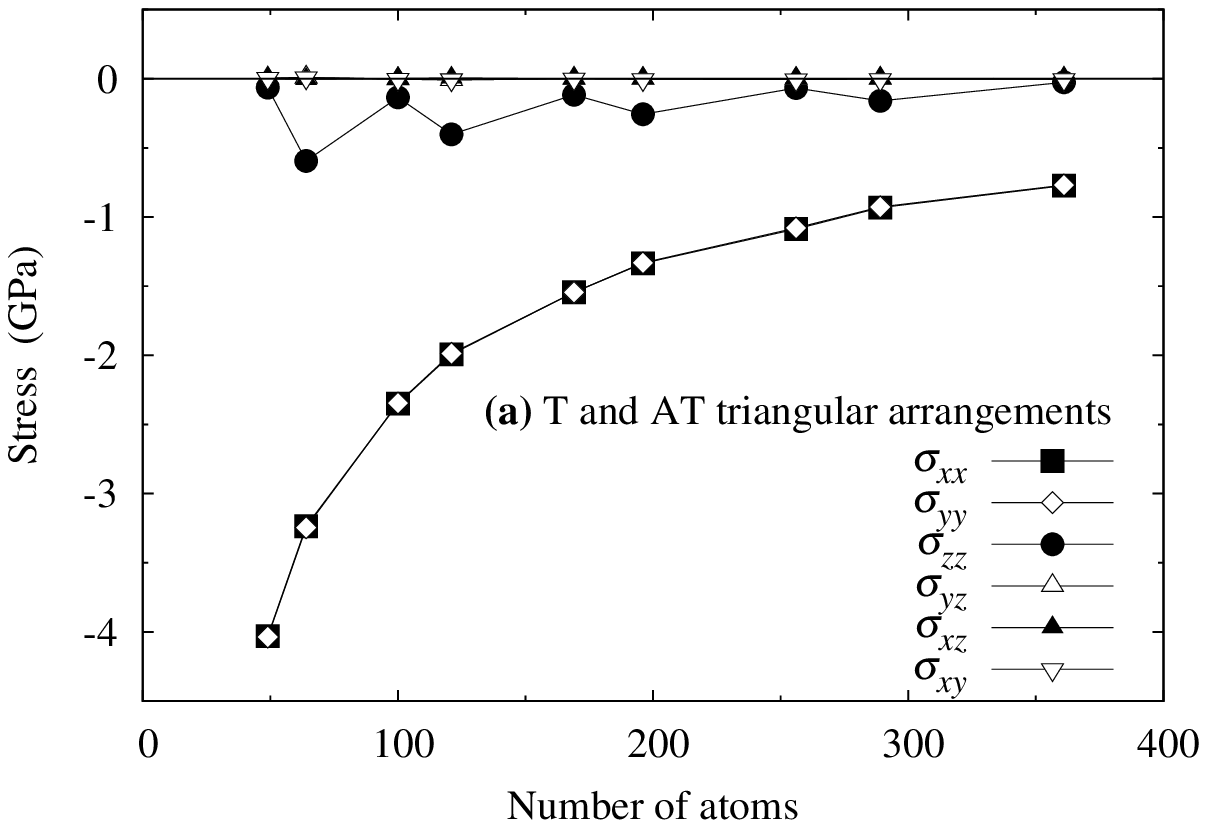}
			\includegraphics[width=0.95\linewidth]{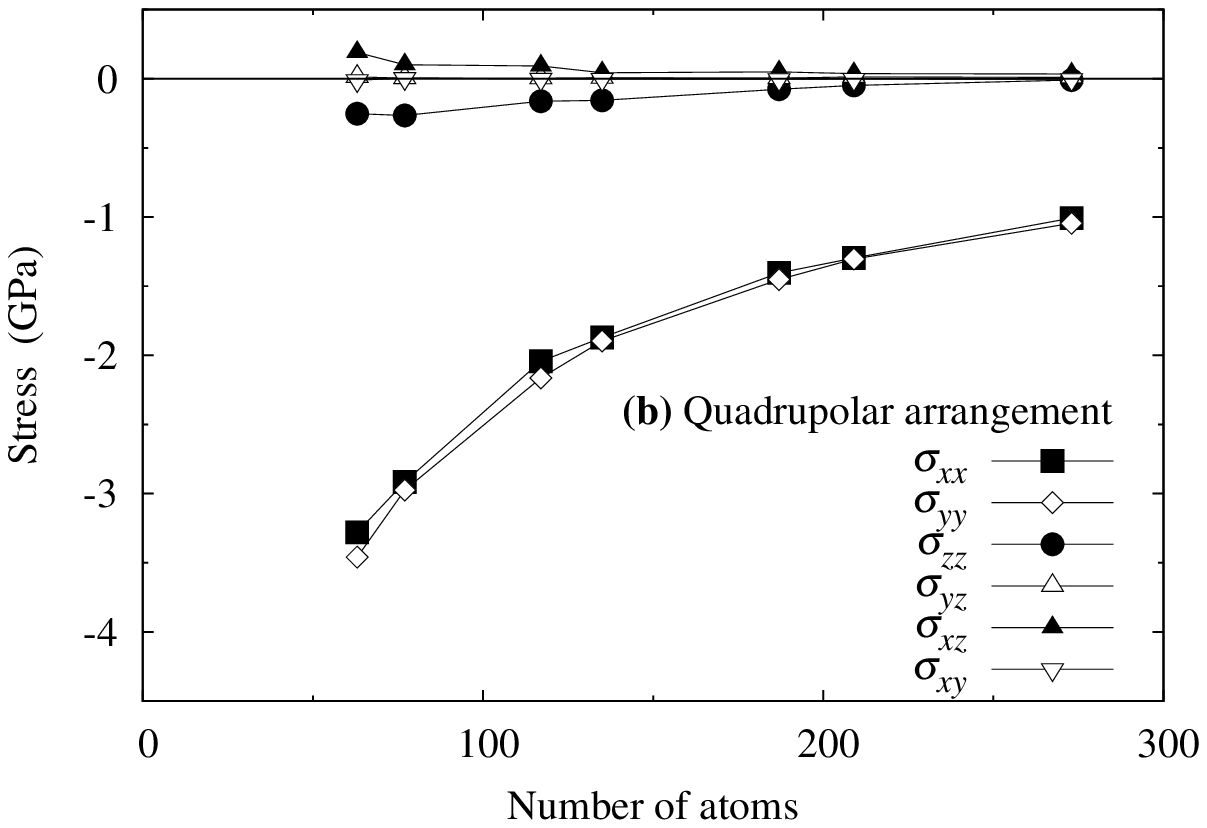}
	\end{center}
	\caption{Homogeneous stress observed in \emph{ab initio} calculations
	for the three periodic dislocation arrangements 
	in the easy core configuration: 
	(a) T and AT triangular; (b) quadrupolar dislocation arrangements.}
	\label{fig:dislo_stress}
\end{figure}

A homogeneous stress is observed in the \emph{ab initio} calculations,
when the periodicity vectors are kept fixed, and when only the atomic positions
are relaxed.
The six components of the corresponding stress tensor are shown 
in Fig.~\ref{fig:dislo_stress} for the three periodic dislocation
arrangements, in the easy core configuration.
The stress components are expressed in the axes $\mathbf{e}_x=\left[ \bar{1}\bar{1}{2} \right]$,
$\mathbf{e}_y=\left[ 1\bar{1}0 \right]$ and $\mathbf{e}_z=\left[ 111 \right]$.
The main components of the stress tensor are $\sigma_{xx}$ and $\sigma_{yy}$,
and the other components can be neglected.
The stress components $\sigma_{xx}$ and $\sigma_{yy}$ vary roughly linearly
with the inverse of the number of atoms, and so with the inverse
of the surface $S$ of the simulation box.
For the two variants of the triangular arrangement, T and AT, the stress components
$\sigma_{xx}$ and $\sigma_{yy}$ are exactly equal [Fig.~\ref{fig:dislo_stress}(a)]. 
Indeed, the threefold symmetry along the $[111]$ direction obeyed by this dislocation
arrangement is also imposed to the homogeneous stress.
On the other hand, the quadrupolar arrangement breaks this symmetry.
As a consequence, $\sigma_{xx}$ and $\sigma_{yy}$ slightly 
differ from each other in this case [Fig.~\ref{fig:dislo_stress}(b)].

The two core fields of the dislocations
composing the dipole are responsible for this homogeneous stress, as shown in Ref.~\onlinecite{CLO09a}.
If $M_{ij}$ is the first moment tensor of the line force distribution
representative of the core field, the homogeneous stress in the simulation 
box is given by \cite{CLO09a}
\begin{equation}
	\sigma_{ij} = -2 \frac{M_{ij}}{S}.
	\label{eq:homogeneous_stress}
\end{equation}
The factor of 2 in this equation arises from the fact that two dislocations
constituting the dipole are introduced within the simulation box.

\begin{figure}[btp]
	\begin{center}
		\includegraphics[width=0.95\linewidth]{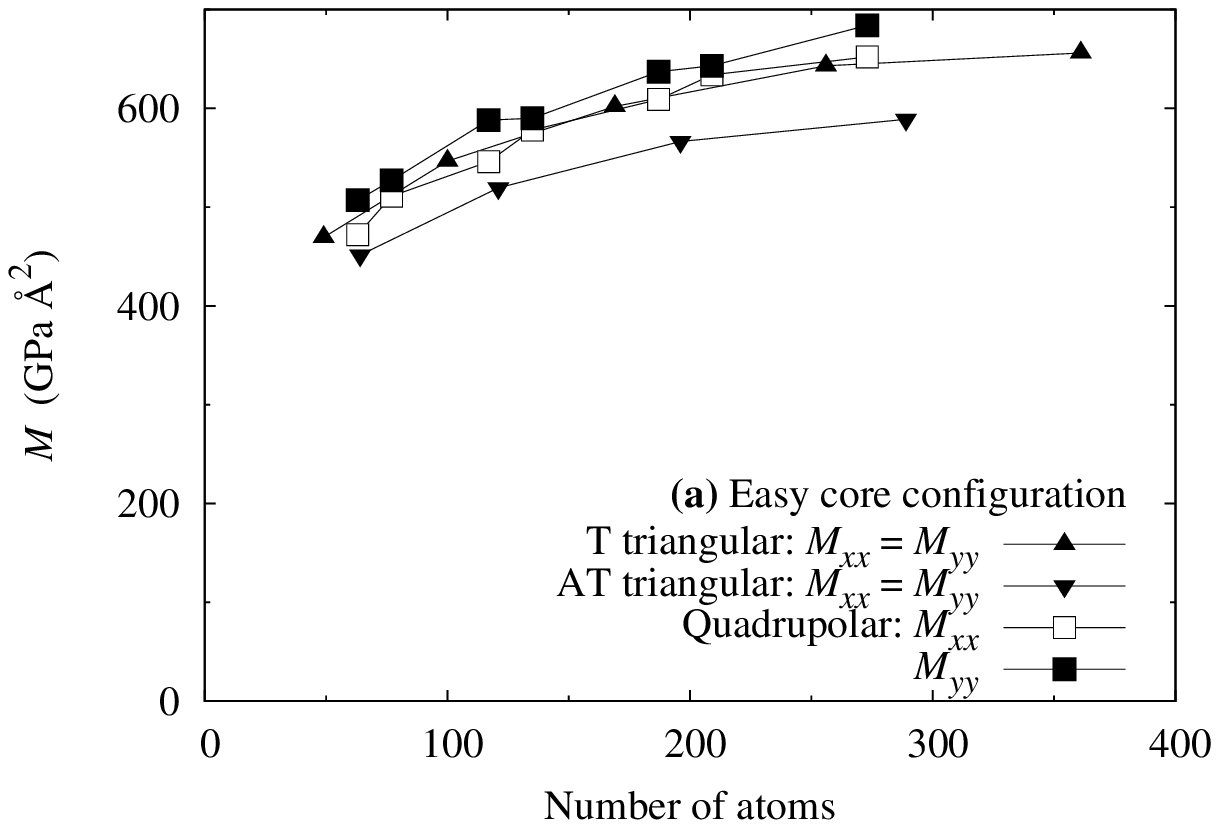}
		\includegraphics[width=0.95\linewidth]{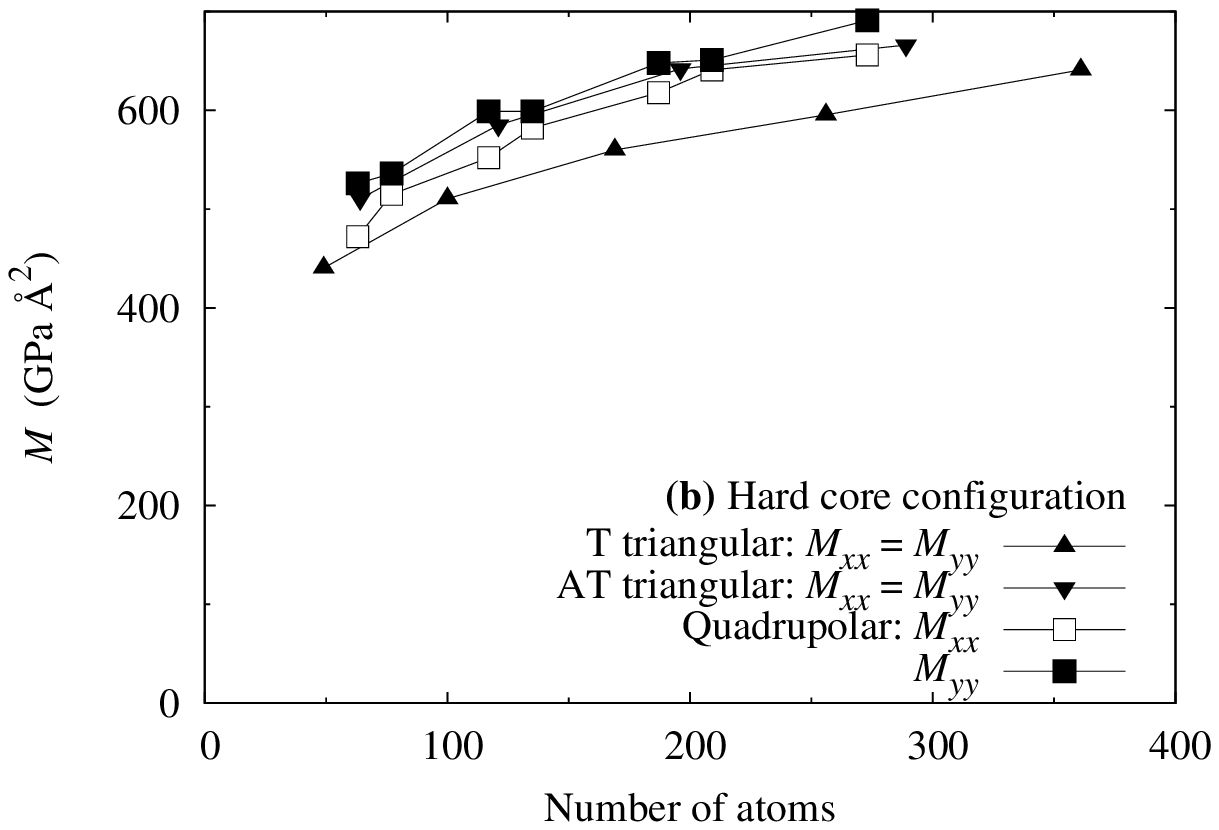}
	\end{center}
	\caption{Moments $M_{xx}$ and $M_{yy}$ of the line force dipoles 
	for the three periodic dislocation arrangements in the 
	(a) easy and (b) hard core configuration.}
	\label{fig:dislo_moments}
\end{figure}

As the stress components other than $\sigma_{xx}$ and $\sigma_{yy}$
can be neglected, the force moment tensor $M$
will have only two non-zero components, $M_{xx}$ and $M_{yy}$, as shown in Eq.~(\ref{eq:homogeneous_stress}). 
These two components deduced from the stress computed from DFT
calculations are represented in Fig.~\ref{fig:dislo_moments}(a) for the easy
core configuration, and in Fig.~\ref{fig:dislo_moments}(b)
for the hard core configuration.
Within the two triangular arrangements, the core field is a pure biaxial 
dilatation ($M_{xx}=M_{yy}$), whereas the core field has a small distortion
component ($M_{xx}-M_{yy}\neq0$) within the quadrupolar arrangement.
This distortion is associated with the broken threefold 
symmetry within the quadrupolar periodic arrangement. It may arise from a polarizability
\cite{DED78b,SCH84,PUL86} of the core field: the moments $M$ characterizing the 
core field may depend on the stress applied to the dislocation core.
Such a polarizability may also be the reason why the moments obtained
within the T variant of the triangular arrangement slightly differ 
from the moments obtained within the AT variant (Fig.~\ref{fig:dislo_moments}).

One can also observe a dependence of the moments with the size 
of the simulation box.
The cell size that can be reached in DFT does not allow us to obtain well-converged values for the 
moments. Nevertheless, the core field of the screw dislocation in iron 
can be considered as a pure biaxial dilatation of amplitude $M_{xx}=M_{yy}=650\pm50$\,GPa\,\AA$^2$.
Interestingly, the easy and the hard core configurations
are characterized by the same moments [Figs.~\ref{fig:dislo_moments}(a) and 
\ref{fig:dislo_moments}(b)], although the atomic structures
of the dislocation cores are completely different between these two configurations.
This probably indicates that the dislocation core field does not
arise from perturbations due to the atomic nature of the core, 
but rather from the anharmonicity of the elastic behavior.

\subsection{Simulations with relaxed periodicity vectors}

\emph{Ab initio} simulations, in which the periodicity vectors
are allowed to relax so as to minimize the energy, have also been performed.
In this case, a homogeneous strain, $\varepsilon^0_{ij}$, can be computed.
This strain is related to the core field of the two dislocations
within the simulation box, through the relation \cite{CLO09a}
\begin{equation*}
	\varepsilon^0_{ij} = 2S_{ijkl}\frac{M_{kl}}{S},
\end{equation*}
where the elastic compliances $S_{ijkl}$ are the inverse 
of the elastic constants\footnote{$S_{ijkl}C_{klmn}=\frac{1}{2}\left( \delta_{im}\delta_{jn}
+ \delta_{in}\delta_{jm}\right)$}.
If the dislocation core field is assumed to be an elliptical 
line source of expansion characterized by the two non-zero
moments $M_{xx}$ and $M_{yy}$, 
the six components of the homogeneous
strain are given by
\begin{equation}
	\begin{split}
	  \varepsilon_{xx}^0 & = \frac{C^{'}_{33}}{C^{'}_{33}(C^{'}_{11}+C^{'}_{12})-2{C^{'}_{13}}^2} \frac{M_{xx}+M_{yy}}{S} \\
		& + \frac{C^{'}_{44}}{C^{'}_{44}(C^{'}_{11}-C^{'}_{12})-2{C^{'}_{15}}^2} \frac{M_{xx}-M_{yy}}{S} ,\\
	\varepsilon_{yy}^0 & = \frac{C^{'}_{33}}{C^{'}_{33}(C^{'}_{11}+C^{'}_{12})-2{C^{'}_{13}}^2} \frac{M_{xx}+M_{yy}}{S} \\
		& - \frac{C^{'}_{44}}{C^{'}_{44}(C^{'}_{11}-C^{'}_{12})-2{C^{'}_{15}}^2} \frac{M_{xx}-M_{yy}}{S} ,\\
	\varepsilon_{zz}^0 & = -\frac{2 C^{'}_{13}}{C^{'}_{33}(C^{'}_{11}+C^{'}_{12})-2{C^{'}_{13}}^2} \frac{M_{xx}+M_{yy}}{S} ,\\
	\varepsilon_{yz}^0 & = 0,\\
	\varepsilon_{xz}^0 & = - \frac{C^{'}_{15}}{C^{'}_{44}(C^{'}_{11}-C^{'}_{12})-2{C^{'}_{15}}^2} \frac{M_{xx}-M_{yy}}{S},\\
	\varepsilon_{xy}^0 & = 0 ,
	\end{split}
	\label{eq:relaxed_strain}
\end{equation}
where the elastic constants $C^{'}_{nm}$ are expressed in Voigt notation, 
in the axes $\mathbf{e}_x=\left[ \bar{1}\bar{1}{2} \right]$,
$\mathbf{e}_y=\left[ 1\bar{1}0 \right]$, and $\mathbf{e}_z=\left[ 111 \right]$.
This shows that, for positive moments $M_{xx}$ and $M_{yy}$, 
a dilatation perpendicular to the dislocation  
and a contraction parallel to the dislocation line are produced.
This exactly corresponds to what we observe in the \emph{ab initio} calculations. 
Thus one can define a dislocation formation volume perpendicular
to the dislocation line, $\delta V_{\bot}=(\varepsilon_{xx}^0+\varepsilon_{yy}^0)S/2$
and a formation volume parallel to the dislocation line, $\delta V_{\parallel}=\varepsilon_{zz}^0S/2$, 
where the values of the formation volume are defined per unit of dislocation line.
The DFT results 
are given in Table \ref{tab:abinitio_relaxed}. 
The expressions of the homogeneous strain [Eq. (\ref{eq:relaxed_strain})] also lead to a ratio between 
the two dislocation formation volumes depending only on the elastic 
constants as
\begin{equation*}
  \frac{\delta V_{\bot}}{\delta V_{\parallel}} = - \frac{C^{'}_{33}}{C^{'}_{13}}.
\end{equation*}
Using the values of the elastic constants calculated in DFT, 
we predict $\delta V_{\bot} / \delta V_{\parallel} = -2.0$.
This is in reasonably good agreement with the values obtained from the homogeneous strain computed in 
\emph{ab initio} calculations (Table~\ref{tab:abinitio_relaxed}).

\begingroup
\squeezetable
\begin{table}[btp]
	\caption{Formation volumes $\delta V_{\bot}$ perpendicular and $\delta V_{\parallel}$ 
	parallel to the dislocation line per unit of dislocation line computed in DFT
	for the three dislocation arrangements. 
	$N$ is the number of atoms contained within the simulation box. 
	$M_{xx}$ and $M_{yy}$ are the moments of the dislocation core field 
	deduced from the components $\varepsilon^0_{xx}$ and $\varepsilon^0_{yy}$
	of the homogeneous strain.}
	\label{tab:abinitio_relaxed}
	\subfigure[Easy core configuration]{
	\begin{ruledtabular}
	\begin{tabular}{lcccccc}
			& $N$	& $\delta V_{\bot}$	& $\delta V_{\parallel}$
			& $\delta V_{\bot} / \delta V_{\parallel}$	
			& $M_{xx}$ 	& $M_{yy}$ 	\\
			& 	& (\AA$^2$)	& (\AA$^2$)
			& 	
			& (GPa\,\AA$^2$)	& (GPa\,\AA$^2$)	\\
		\hline
		T tri.	& 169	& 4.0	& --1.5	& --2.7	& 551	& 545	\\
		AT tri.	& 121	& 3.4	& --1.2	& --2.8	& 463	& 463	\\
		AT tri.	& 196	& 3.9	& --1.2	& --3.1	& 530	& 545	\\
		quadru.	& 135	& 3.9	& --1.4	& --2.8	& 527	& 541	\\
	\end{tabular}
	\end{ruledtabular}}
	\subfigure[Hard core configuration]{
	\begin{ruledtabular}
	\begin{tabular}{lcccccc}
			& $N$	& $\delta V_{\bot}$	& $\delta V_{\parallel}$
			& $\delta V_{\bot} / \delta V_{\parallel}$	
			& $M_{xx}$ 	& $M_{yy}$ 	\\
			& 	& (\AA$^2$)	& (\AA$^2$)
			& 	
			& (GPa\,\AA$^2$)	& (GPa\,\AA$^2$)	\\
		\hline
		T tri.	& 169	& 3.5	& --1.3	& --2.6	& 488	& 465	\\
		AT tri.	& 121	& 3.9	& --1.3	& --3.0	& 531	& 531	\\
		AT tri.	& 196	& 4.6	& --1.6	& --2.9	& 646	& 619	\\
		quadru.	& 135	& 3.7	& --1.3	& --2.7	& 518	& 493	\\
	\end{tabular}
	\end{ruledtabular}}
\end{table}
\endgroup

The moments $M_{xx}$ and $M_{yy}$ characterizing the dislocation core field
can be deduced from the homogeneous strain computed in DFT, 
using the system of equations (\ref{eq:relaxed_strain}).
We choose to derive $M_{xx}$ and $M_{yy}$ from the components $\varepsilon_{xx}$
and $\varepsilon_{yy}$ of the strain, and the resulting values 
are given in Table~\ref{tab:abinitio_relaxed} for
the three arrangements.
These values are in good agreement with those derived from atomistic simulations with 
fixed periodicity vectors (Sec. \ref{sec:fixed_periodicity_vectors}).
For all periodic arrangements of dislocations, the dislocation core field 
can be considered as a pure biaxial dilatation, 
and we neglect the difference between $M_{xx}$ and $M_{yy}$.

\section{Core field in atomistic simulations}

Now that the dislocation core field has been characterized, 
we examine its influence on atomistic simulations. 
First we look at the atomic displacements observed in \emph{ab initio}
calculations: part of this displacement arises from the core field.
Then we show that the dislocation core energies
can be extracted from these calculations when the core field contribution is considered
in the elastic energy.

\subsection{Atomic displacement}

\begin{figure*}[btp]
	\begin{center}
		\subfigure[T triangular periodic arrangement (361 atoms).]{
		\includegraphics[width=0.99\linewidth]{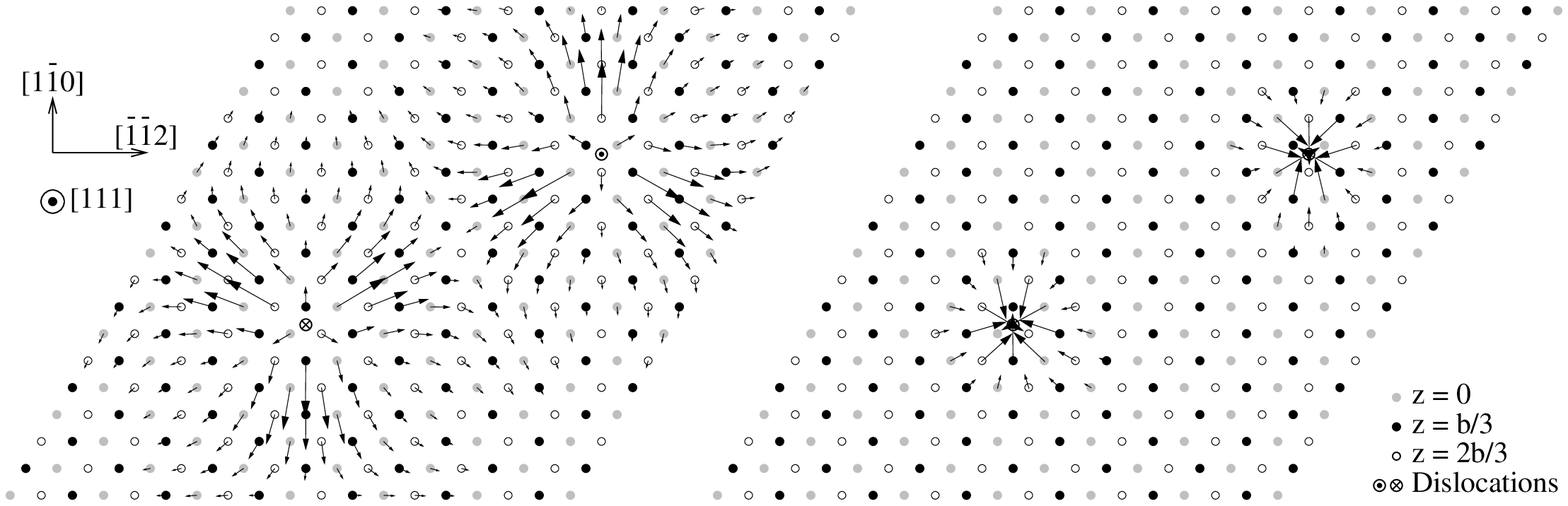}}
		\subfigure[AT triangular periodic arrangement (289 atoms).]{
		\includegraphics[width=0.99\linewidth]{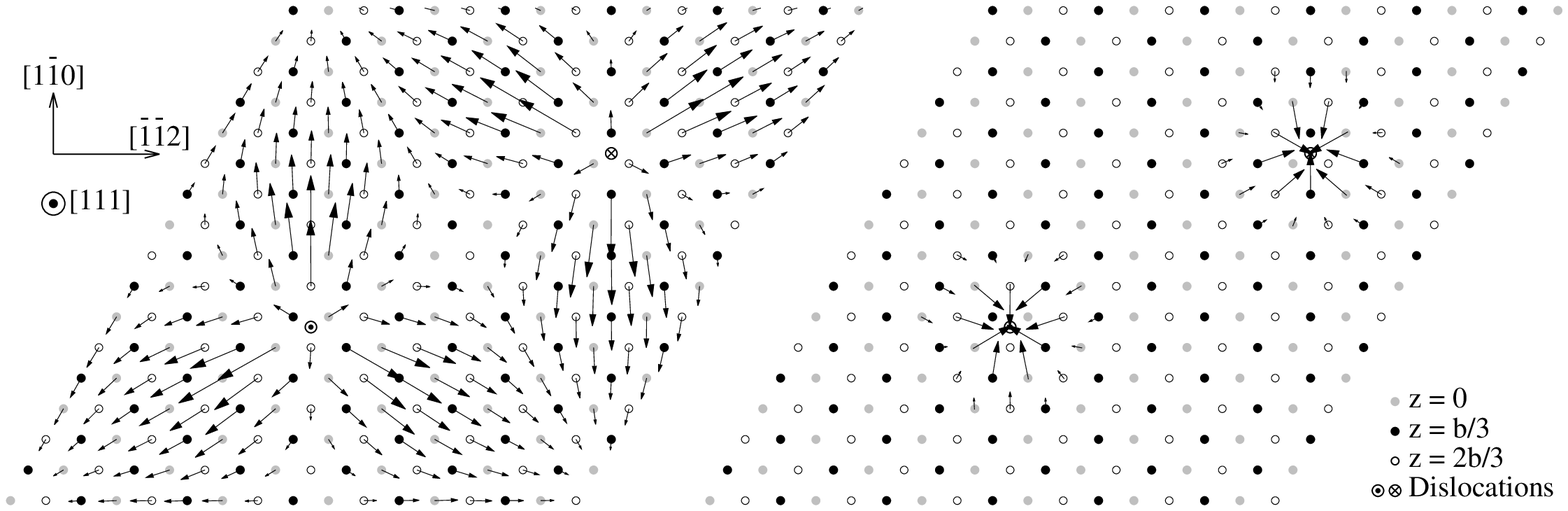}}
		\subfigure[Quadrupolar periodic arrangement (209 atoms).]{
		\includegraphics[width=0.99\linewidth]{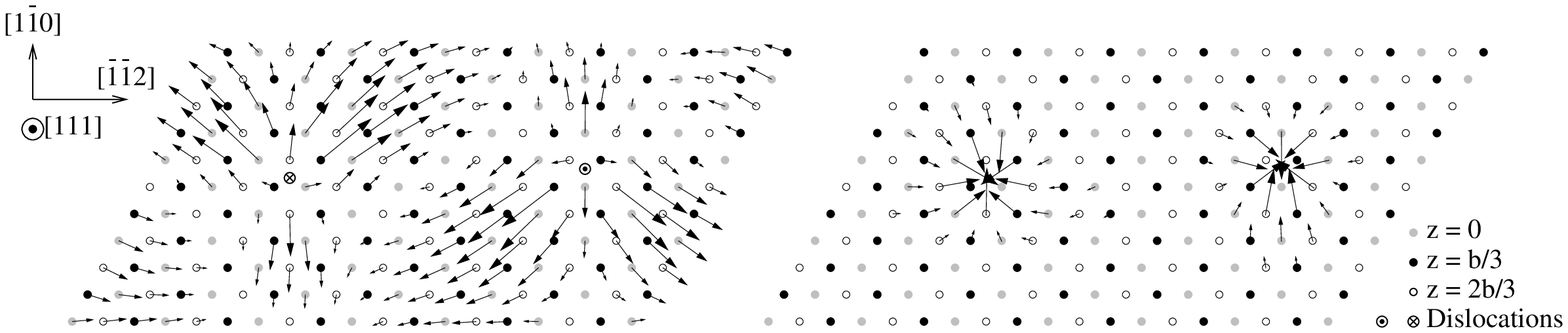}}
	\end{center}
	\caption{Planar displacement map of a periodic unit cell containing a screw dislocation dipole
	in the easy core configuration,
	as obtained from \emph{ab initio} calculations (left), and after subtraction of the Volterra 
	and the core elastic fields (right). 
	Vectors correspond to the $(111)$ in-plane displacement and have been magnified by a factor of 50.
	Displacements smaller than 0.01\,{\AA} are omitted. 
	For clarity, displacements of the six atoms belonging to the dislocation cores
	are not shown on right panel.}
	\label{fig:displacement_map}
\end{figure*}

Relaxation of atomic positions in \emph{ab initio} calculations
leads to the definition of atomic displacements induced by the periodic array
of dislocation dipole.
The displacement along the $[111]$ direction, 
\ie along the dislocation line, also called the screw component,
is very close to the anisotropic elastic solution 
corresponding to the Volterra field.
There is no substantial contribution of the core field
on this displacement component.

A displacement perpendicular to the dislocation line can also be evidenced
in atomistic simulations, \ie an edge component
(Fig.~\ref{fig:displacement_map}). 
Part of this displacement component corresponds to the
dislocation Volterra field and arises from elastic anisotropy.
The dislocation core field also contributes to the edge component.
The Volterra contribution is more long-ranged than the core field contribution,
as the former varies with the logarithm of the distance 
to the dislocation, whereas the latter varies with the inverse
of this distance. 
Nevertheless, both contributions evidence a similar amplitude in the simulations,
because of the reduced cell size used in DFT calculations.

We can subtract from the atomic displacement given by \emph{ab initio} calculations,
the displacement corresponding to the superposition of the Volterra and the dislocation core fields, 
as predicted by the anisotropic linear elasticity taking full account
of the periodic boundary conditions \cite{CAI03}.
The resulting map, represented in Fig.~\ref{fig:displacement_map},
shows that the elastic modeling manages to reproduce 
the \emph{ab initio} displacements for all periodic arrangements
even close to the dislocation core.
The anisotropic linear elasticity fails to reproduce the \emph{ab initio} atomic displacement
only on the atoms within the core: displacements in the core are too large
for applying a small perturbation theory such as elasticity.

\subsection{Dislocation core energy}

\begin{figure}[btp]
	\begin{center}
		\includegraphics[width=0.99\linewidth]{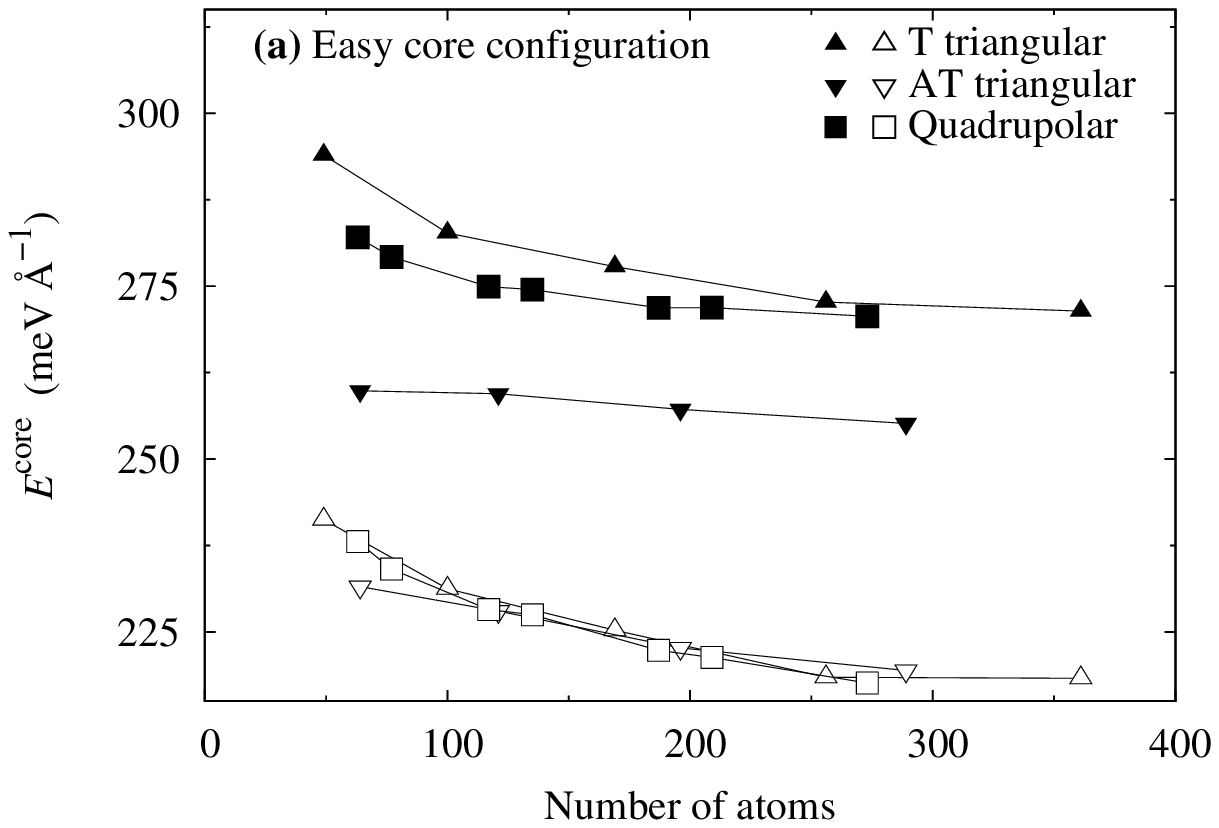}
		\includegraphics[width=0.99\linewidth]{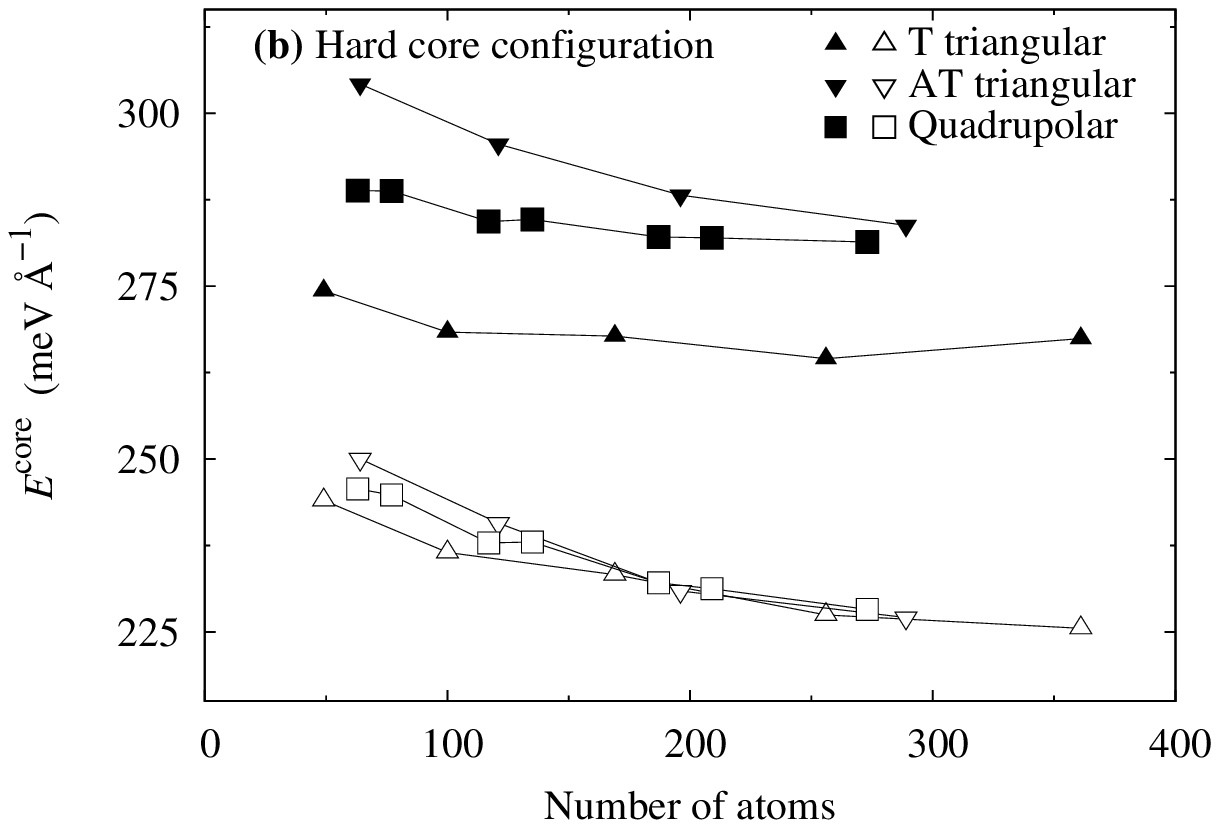}
	\end{center}
	\caption{Core energy of the screw dislocation in the (a) easy
	and the (b) hard core configuration. Solid symbols correspond to the core energies
	obtained when only the Volterra field is considered, 
	and open symbols, to the core energies when both the Volterra and the core fields
	are taken into account ($r_{\rm c}=3$\,\AA).}
	\label{fig:dislo_core}
\end{figure}

The screw dislocation core energy is deduced from DFT simulations 
by subtracting the elastic energy to the excess energy of the unit cell computed 
by \emph{ab initio} techniques. 
The elastic energy is calculated by taking into account the elastic anisotropy
and the periodic boundary conditions \cite{CAI03}.

If only the Volterra contribution is considered in the elastic energy,
the resulting core energies strongly depend on the periodic arrangement
(Fig.~\ref{fig:dislo_core}). As shown in Ref.~\onlinecite{CLO09a}, one 
can not conclude on the relative stability of the two different
core configurations of the screw dislocation. The AT triangular geometry
and the quadrupolar geometry predict that the easy core 
is the most stable configuration, whereas the T triangular arrangement
leads to the opposite conclusion.
Then if both the Volterra and the core fields are considered
in the elastic energy, the core energies do not depend anymore
on the periodic arrangement.
A cell size dependence has been evidenced
(Fig.~\ref{fig:dislo_core}), and 
in all geometries the easy core is more stable than the hard
core configuration.
The convergence is reached for a reasonable number of atoms: 
$E^{\rm core}=219\pm1$\,meV\,\AA$^{-1}$ for the easy core configuration,
and $227\pm1$\,meV\,\AA$^{-1}$ for the hard core configuration.
These core energies are given for a core radius 
$r_{\rm c}=3$\,\AA, as this value of $r_{\rm c}$ has been found to lead 
to a reasonable convergence of the core energy with respect to the cell size
(\cf Appendix \ref{sec:Ecore_rc}).

To better understand how the core energy converges, one can decompose
the elastic energy into the different contributions,
and look how these contributions vary with the length scale. 
If one neglects the dislocation core field, the elastic energy
of the simulation box containing a dislocation dipole is given by
\begin{equation}
	E^{\rm elas}_{\rm V} = 2E^{\rm elas}_{\rm c}
		+ b_i K^0_{ij}b_j \ln{\left( A/r_{\rm c} \right)}
		+ E^{\rm inter}_{\rm V-V} ,
	\label{eq:Eelas_V}
\end{equation}
where $K^0$ is a second rank tensor, which depends only on the 
elastic constants, and $r_{\rm c}$ is the dislocation core cutoff. 
The two first terms on the right hand side define the elastic energy 
of the dipole contained in the simulation box: $E^{\rm elas}_{\rm c}$
is the core traction contribution \cite{CLO09b}, and the second 
term corresponds to the cut contribution.
$E^{\rm inter}_{\rm V-V}$ corresponds to the interaction 
of the dislocation dipole with its periodic images \cite{CAI03}.
If the core field is taken into account \cite{CLO11a},
the elastic energy becomes
\begin{equation}
	E^{\rm elas} = E^{\rm elas}_{\rm V} 
		+ M_{ij}K^2_{ijkl}M_{kl}\frac{1}{{r_{\rm c}}^2}
		+ 2 E^{\rm inter}_{\rm V-c} 
		+  E^{\rm inter}_{\rm c-c},
	\label{eq:Eelas_V_core}
\end{equation}
where the fourth rank tensor $K^2$, which only depends on the elastic constants,
enables one to calculate the core field contribution 
to the dislocation self energy.
$E^{\rm inter}_{\rm V-c}$ and $E^{\rm inter}_{\rm c-c}$
correspond to the interaction of the dislocation core field 
with respectively the Volterra field, and the core field of the other dislocations, 
\ie the second dislocation composing the dipole, as well as 
the image dislocations due to the periodic boundary conditions.
When the periodicity vectors, $\mathbf{u}$ and $\mathbf{v}$,
and the dipole cut, $\mathbf{A}$, are scaled by the same factor $\lambda$,
$E^{\rm inter}_{\rm V-c}$ varies as $1/\lambda$, 
and $E^{\rm inter}_{\rm c-c}$, as $1/\lambda^2$.
The number $N$ of atoms in the simulation box is proportional 
to $\lambda^2$.
The comparison of equations (\ref{eq:Eelas_V}) and (\ref{eq:Eelas_V_core})
shows then that the neglect of the core field
leads to a core energy that converges as $N^{-1/2}$.
In the case of dislocation periodic arrangements, which are centrosymmetric 
like the quadrupolar arrangement of Fig.~\ref{fig:PBC}(c), 
$E^{\rm inter}_{\rm V-c}$ vanishes in Eq.~(\ref{eq:Eelas_V_core}).
The core energy converges thus as $N^{-1}$ when the core field is neglected.
In all cases, this convergence is too slow to extract
a meaningful core energy from \emph{ab initio} calculations.
Moreover, the obtained value does not really correspond to the 
core energy $E^{\rm core}$: it results from the summation of $E^{\rm core}$
and the core field self elastic energy,
$M_{ij}K^2_{ijkl}M_{kl} 1 / {r_{\rm c}}^2$.

Another interesting feature comes from the linear dependency of $E^{\rm inter}_{\rm V-c}$
with the Burgers vector. 
Going from the easy to the hard core configuration of the 
dislocation dipole, \ie inverting the sign of the Burgers 
vector, one only changes the sign of $E^{\rm inter}_{\rm V-c}$,
whereas all other contributions to the elastic energy 
remain constant. The easy and the hard core
configurations have indeed a core field with the same amplitude 
(Fig.~\ref{fig:dislo_moments}). 
The contribution $E^{\rm inter}_{\rm V-c}$ is positive for 
a dislocation dipole in its easy core configuration 
within the T triangular arrangement [Fig.~\ref{fig:PBC}(a)]. 
Therefore, when the core field contribution is not included
into the elastic energy, one underestimates the stability of the 
easy core configuration with respect to the hard core configuration
within this geometry. 
The AT triangular arrangement leads to the opposite conclusion
[Fig.~\ref{fig:PBC}(b)], since $E^{\rm inter}_{\rm V-c}$ becomes
negative for the easy core configuration.

The underestimation or overestimation of the stability of the easy core
illustrates the importance of considering the dislocation
core field in the elastic energy, when extracting quantitative 
properties from atomistic simulations.
This is especially true for \emph{ab initio} calculations,
 in which the small cell size makes it difficult to obtain
converged values.
Such a conclusion is not restricted to the calculation 
of dislocation core energies.
For instance, extraction from atomistic simulations
of the Peierls energy barriers, and of the associated Peierls stresses,
will also require the complete modeling of the dislocation 
core field.

\subsection{Dislocation line energy and line tension}

It is interesting to evaluate the different contributions 
to the dislocation line energy. Using the developed elastic model, 
the energy
\footnote{The different energy contributions have been calculated 
for a core radius $r_{\rm c}=3$\,{\AA}.}
of a straight screw dislocation contained in a cylinder
of radius $R_{\infty}$ is
\begin{multline}
	E = E^{\rm core} + E_{\rm c}^{\rm elas} 
	+ \frac{1}{2} b_i K^0_{ij} b_j \ln{\left( \frac{R_{\infty}}{r_{\rm c}}\right)} \\
	+ \frac{1}{2} M_{ij} K^2_{ijkl} M_{kl} \frac{1}{ {r_{\rm c}}^2}.
	\label{eq:Edislo}
\end{multline}
The core energy has been found to be $E^{\rm core} = 219 \pm 1$\,meV\,\AA$^{-1}$ for the 
easy core configuration, and $227 \pm 1$\,meV\,\AA$^{-1}$ for the hard 
core configuration, with $r_{\rm c}=3$\,{\AA}.
The Volterra elastic field leads to two energy contributions: 
the contribution of the core traction \cite{CLO09b} is 
$E_{\rm c}^{\rm elas} = 1$\,meV\,\AA$^{-1}$, for a dislocation cut
corresponding to a $\{110\}$ glide plane,
and the cut contribution, which corresponds to the third term in Eq.~(\ref{eq:Edislo}),
is equal to 1.60\,eV\,\AA$^{-1}$, where we have assumed the ratio
$R_{\infty}/r_{\rm c}=10^4$, which corresponds to a characteristic
dislocation density of $10^{11}$\,m$^{-2}$.
Finally, the core field contribution to the elastic energy, 
which corresponds to the last term in Eq.~(\ref{eq:Edislo}),
is equal to 42\,meV\,\AA$^{-1}$.
It is clear that most of the dislocation energy arises from the 
Volterra elastic field, and is associated with the cut contribution. 
Other contributions, which are all associated with 
the dislocation core, account for about 14\,\% of the dislocation
energy. In particular, the contribution of the core field 
is less than 3\,\%.

\begin{figure}[]
  \begin{center}
    \includegraphics[width=0.99\linewidth]{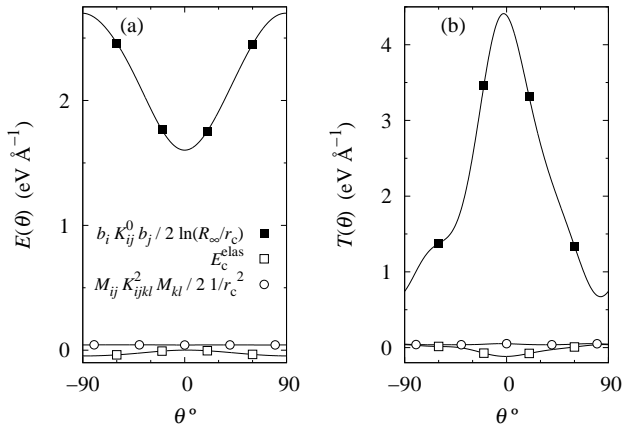}
  \end{center}
  \caption{Elastic contributions [Eq.~(\ref{eq:Edislo})]
  to the dislocation (a) line energy $E(\theta)$ 
  and (b) line tension $T(\theta) = E(\theta) + {\md^2 E(\theta)}/{\md\theta^2}$
  as a function of the dislocation character $\theta$ in the $\left\{ 110 \right\}$
  glide plane.
  $\theta=0$ for the screw orientation.}
  \label{fig:lineTension}
\end{figure}

The dislocation line tension is actually more important than the line energy
as it controls the shape of dislocation loops and curved dislocations\cite{WIT59}. 
It is defined as 
\begin{equation*}
  T(\theta) = E(\theta) + \frac{\md^2 E(\theta)}{\md \theta^2},
\end{equation*}
if $E(\theta)$ is the dislocation line energy [Eq.~(\ref{eq:Edislo})]
written as a function of the dislocation character $\theta$, \ie the dislocation orientation. 
We can evaluate this line tension by considering all elastic contributions
entering in the dislocation line energy and neglecting the dependence
of the dislocation core energy with $\theta$ which we do not know. 
For the dislocation core field, we assume that the dipole tensor $M_{ij}$
does not depend on the dislocation orientation as we do not have
any information on such a variation:
the variation with the dislocation orientation of the associated line energy 
only arises from elastic anisotropy.
The different elastic contributions to the dislocation line energy $E(\theta)$
and the associated line tension $T(\theta)$ are shown in Fig.~\ref{fig:lineTension}
for a dislocation character going from an edge orientation ($\theta=\pm90^{\circ}$)
to a screw orientation ($\theta=0$).
The most important contribution to the line tension arises, once again, 
from the cut contribution of the Volterra elastic field.
In view of the values obtained, it looks reasonable to neglect
other elastic contributions, as usually done in line tension models\cite{WIT59}.
Nevertheless, some other studies have obtained higher 
relative contributions of the core field \cite{GEH72,HOA76}, 
which may therefore have a more important effect on the line tension. 

\section{Dislocation interaction with a carbon atom}

We examine in this part how the dislocation core field influences 
the dislocation properties by modifying the way a dislocation
can interact with its environment. 
First, we study the interaction between a $\langle111\rangle$
screw dislocation and a carbon atom in $\alpha$-iron. 

In Ref.~\onlinecite{CLO08}, the binding energy between a carbon atom
and a screw dislocation in iron 
as predicted by the linear anisotropic elasticity has been compared to the one 
calculated by atomistic simulations based on an empirical potential approach \cite{BEC07}.
A quantitative agreement between both modeling techniques
was obtained as long as the C atom 
was located at a distance greater than 2\,{\AA} from the dislocation core. 
This result evidences the ability of linear elasticity to predict this interaction.
The empirical potential for iron \cite{MEN03,ACK04} used in Ref.~\onlinecite{CLO08}
does not lead to any core field for the screw dislocation,
at variance with the present \emph{ab initio} results.
It is interesting to include now the core field contribution into the elastic field
of the screw dislocation, to investigate its influence on the interaction 
between a carbon atom and a screw dislocation.
Such a contribution to the interaction energy between a solute atom and a dislocation
has already been considered in the case of a substitutional impurity by Fleischer\cite{FLE63} 
who showed that it partly contributes to the solid solution hardening.

\subsection{Carbon atom description}

First we need to deduce from \emph{ab initio} calculations
a quantitative representation of a carbon atom embedded in an iron matrix. 
A solute atom is modeled in elasticity theory by its dipolar tensor, $P_{ij}$,
which corresponds to the first moments of the equilibrated point force 
distribution equivalent to the impurity.
This tensor is deduced from atomistic simulations of one solute
atom embedded in the solvent (\cf Appendix \ref{sec:C}). 

Carbon atoms are found in the octahedral interstitial sites of the bcc lattice.
The dipolar tensor, $P_{ij}$, expressed in the cubic axes, is diagonal
with only two independent components, because of the tetragonal symmetry 
of the octahedral site. Three variants can be obtained, depending on the orientation 
of the tetragonal symmetry axis. 
For the $\left[ 001 \right]$ variant of the C atom,
\emph{ab initio} calculations lead to $P_{xx}=P_{yy}=8.9$ and $P_{zz}=17.5$\,eV
(\cf Appendix \ref{sec:C}). 

\subsection{Binding energy}

\begin{figure}[btp]
	\begin{center}
		\includegraphics[width=0.99\linewidth]{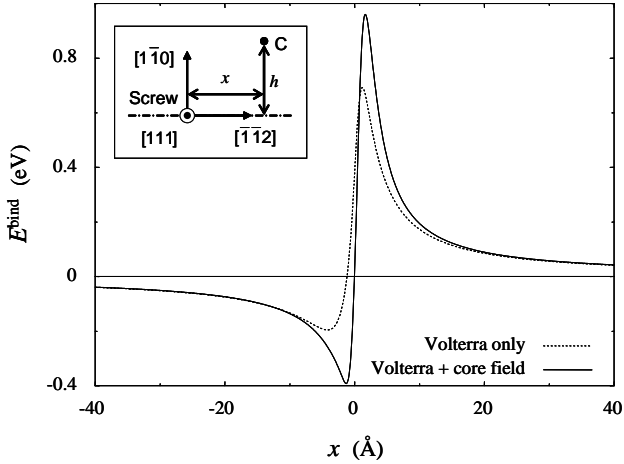}
	\end{center}
	\caption{Binding energy between a $1/2[111](1\bar{1}0)$
	screw dislocation and a carbon atom for different positions $x$
	of the dislocation in its $(1\bar{1}0)$ glide plane. The C atom lies in a [100]
	octahedral site in the plane $h=d_{110}\approx2.04$\,{\AA} 
	above the glide plane. 
	The binding energy is calculated using the anisotropic elasticity theory
	considering only the Volterra field, 
	or both the Volterra and the core fields of the dislocation.}
	\label{fig:C_binding}
\end{figure}

Linear elasticity theory predicts that the binding energy between the carbon atom 
characterized by its dipolar tensor, $P_{ij}$, and the screw dislocation is given by
\begin{equation}
	E^{\rm bind} = P_{ij}\varepsilon_{ij}^{\rm d},
	\label{eq:C_binding}
\end{equation}
where $\varepsilon_{ij}^{\rm d}$ is the elastic strain created by the dislocation.
We use the linear anisotropic elasticity to calculate $\varepsilon_{ij}^{\rm d}$ 
by taking into account only the Volterra field, or both the Volterra
and the core fields created by the dislocation.
We consider that the core field is created by the line force moments 
$M_{xx}=M_{yy}=650$\,GPa\,\AA$^2$ previously deduced.
In Fig.~\ref{fig:C_binding}, we represent the variation of the binding energy 
when the dislocation glides in a $\left\{110\right\}$ plane,
while the carbon atom remains at a fixed distance $h$ from the glide plane.
The first derivative of the plotted function gives the force exerted 
by the C atom on the gliding dislocation.
When the C atom is close enough to the dislocation, the core field
modifies the binding energy. 
In particular, the binding of the C atom is stronger in the attractive region
when the dislocation core field is considered. Thus the pinning of the screw dislocation
by the C atom is enhanced by its core field.
Conversely, the dislocation core field leads to a stronger repulsion in the repulsive region.

When the separation distance between the C atom and the screw dislocation is high enough
($\gtrsim 20$\,\AA), the dislocation core field does not affect anymore 
the binding energy. One can consider that the C atom interacts only 
with the dislocation Volterra field.

\section{Passing properties of a screw dislocation dipole}

We look in this part at how the dislocation core field modifies the equilibrium
properties of a screw dislocation dipole.
Dislocation dipoles play a significant role in single slip straining,
where they can control the material flow stress. Such a situation arises 
for instance in fatigued metals, where dislocations are constrained to glide
in the channels between dislocation walls \cite{MUG05,BRO06b}. 
The saturation stress of the persistent slip bands is then partly controlled
by the critical stress needed to destroy the dislocation dipoles.

\begin{figure}[btp]
	\begin{center}
		\includegraphics[width=0.7\linewidth]{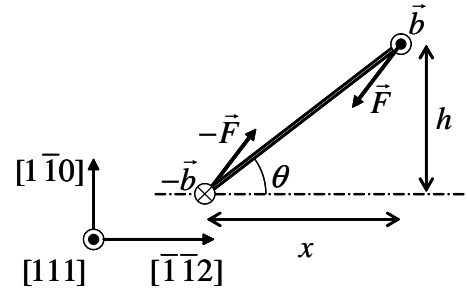}
	\end{center}
	\caption{Screw dislocation dipole. $\mathbf{F}$ is the force exerted by
	one dislocation on the other.}
	\label{fig:dipole_sketch}
\end{figure}

\begin{figure}[btp]
	\begin{center}
		\includegraphics[width=0.99\linewidth]{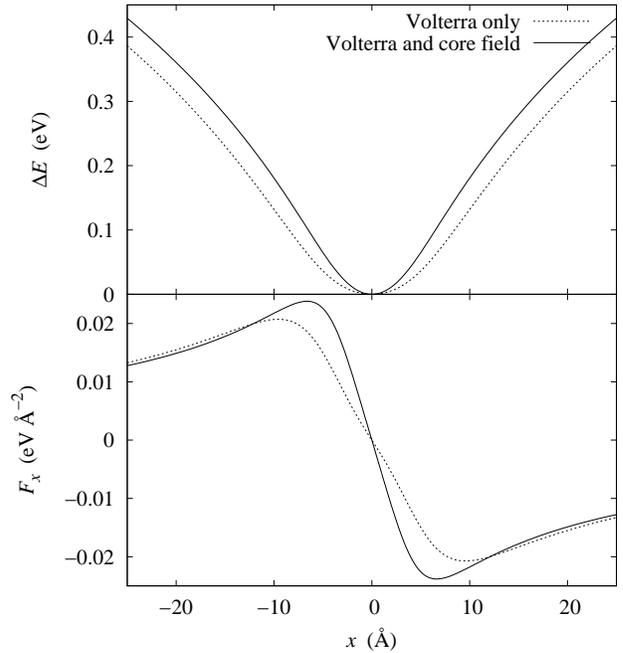}
	\end{center}
	\caption{Variation of the elastic energy, $\Delta E$, 
	and of the $x$ component of the force $\mathbf{F}$ acting
	on the dislocations, $F_x$, with the distance between 
	both dislocations, $x$, for a screw dipole of height $h=10$\,\AA.}
	\label{fig:dipole_h10}
\end{figure}

We consider a screw dislocation dipole in bcc iron.
The dipole is characterized by the height between 
each dislocation glide plane, $h$, and by the projection  
of the dipole vector on the glide plane, $x$, as sketched in Fig.~\ref{fig:dipole_sketch}.
Then we calculate the variation of the interaction energy 
between the two dislocations composing the dipole, $\Delta E$, when dislocations
glide, \ie $h$ is kept fixed while $x$ varies.
This dislocation interaction energy is computed using the linear anisotropic
elasticity \cite{CLO11a}, and considering that the two dislocations 
composing the dipole interact only through the Volterra field,
or through both the Volterra and the core fields.
This variation of energy, $\Delta E$, is represented in Fig.~\ref{fig:dipole_h10} 
for a dipole height $h=10$\,\AA.
The dipole equilibrium angle corresponds to the minimum of $\Delta E$,
\ie $\theta=0$. This value predicted by the anisotropic elasticity is 
equal to the one given by the isotropic elasticity. This contrasts with 
what is found in fcc metals, where the dipole equilibrium angle
strongly depends on elastic anisotropy for a screw dislocation dipole \cite{VEY07}.
The dislocation core field does not modify the dipole equilibrium angle
(Fig.~\ref{fig:dipole_h10}).
Nevertheless, when both the Volterra and the core fields are included 
into the computation of the dipole elastic energy, the energy that defines 
the dipole equilibrium becomes steeper than when the dislocation core field is omitted.
Thus the attraction between both dislocations is stronger when the core field
is taken into account.
This is obvious when looking at the glide component 
of the force exerted by one dislocation on the other one, $F_x$ (Fig.~\ref{fig:dipole_h10}). 
This force is the first derivative of $\Delta E$ with respect to $x$.
When the dislocation core field is included in the dislocation interaction, 
this force goes to a higher maximum value than when only the Volterra elastic field 
is considered. 
To destroy the dislocation dipole, one needs to exert on a dislocation
a force that is higher than the force arising from the interaction with the other dislocation. 
This shows that the dislocation core field leads to a more stable dipole.

\begin{figure}[btp]
	\begin{center}
		\includegraphics[width=0.99\linewidth]{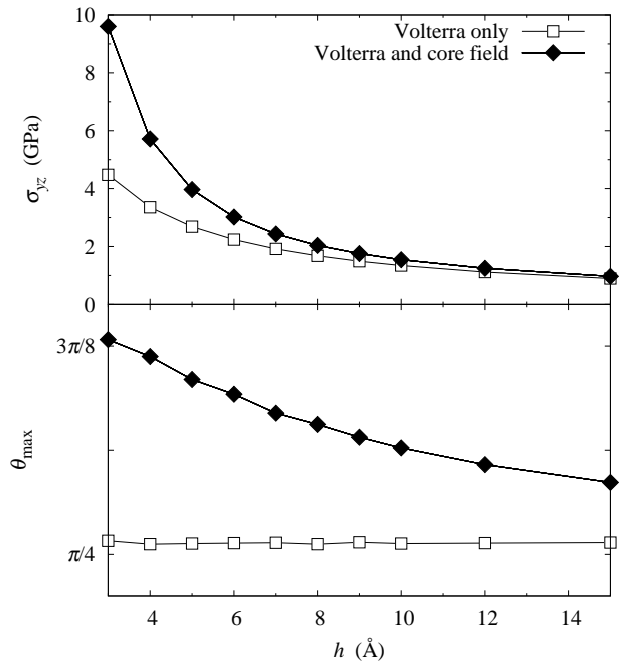}
	\end{center}
	\caption{Variation of the passing stress, $\sigma_{yz}$, and of the passing
	angle, $\theta_{\rm max}$, with the dislocation dipole height, $h$.}
	\label{fig:dipole_passing}
\end{figure}

If a homogeneous stress is applied, the gliding force on each dislocation
is simply the Peach-Koehler force, $b\sigma_{yz}$.
As the applied stress is homogeneous, no force originates from 
the dislocation core field \cite{CLO11a}. Therefore the dipole passing stress, \ie 
the applied stress needed to destroy the dislocation dipole, 
is given by the maximum of the glide component $F_x$ of the Peach-Koehler force
divided by the norm of the Burgers vector. This passing stress
depends on the dipole height, $h$, as shown in Fig.~\ref{fig:dipole_passing}.
The inclusion of the core field into the dislocation interaction 
leads to a higher passing stress, especially for small dipole heights. 
But the effect is relevant at a spacing where the dipole would
certainly have cross-slipped to annihilation.
For large dipole heights ($h\geq20$\,\AA), the dislocation core field does not influence
too much the passing stress, and one can consider that dislocations
interact only through the Volterra elastic field to calculate 
the passing stress.

Without the dislocation core field, the dipole passing angle, $\theta_{\rm max}$, 
does not depend on the dipole height (Fig.~\ref{fig:dipole_passing}). 
The value given by the anisotropic elasticity is close to the $\pi/4$
value predicted by the isotropic elasticity. 
The core field leads to a passing angle, which depends on the dipole height,
and which strongly deviates from $\pi/4$ for small dipole heights.
Such a dependence of the passing angle with the dipole height 
has also been obtained by Henager and Hoagland 
for edge dislocation dipoles in fcc metals \cite{HEN04,HEN05}.
They obtained a stronger influence of the core field
on the dislocation interaction than in the present study. 
In their case, the core field contribution
can be neglected only when the two dislocations are separated by more than $50b$, 
\ie more than 10\,nm.

\section{Conclusions}

The approach developed in the preceding paper to model 
dislocation core field \cite{CLO11a}, 
has been applied here to the $\langle111\rangle$ screw dislocation 
in $\alpha$-iron.
Using \emph{ab initio} calculations, we have shown that a screw 
dislocation creates a core field corresponding to a dilatation
perpendicular to the dislocation line. 
The core field modeling within the anisotropic linear elasticity
perfectly reproduces the atomic displacements observed in \emph{ab initio}
calculations. 
It also allows to derive from atomistic simulations converged values
of the dislocation core energies.
The developed approach illustrates the necessity to consider the
dislocation core field when extracting quantitative information from 
atomistic simulations of dislocations.

Then the elastic modeling of the screw dislocation has been used 
to study the interaction energy between the dislocation and 
a carbon atom. The dislocation core field increases the binding 
of the C atom when both defects are close enough (less than $\sim$20\,\AA).
At larger separation distances, the C atom interacts only with 
the dislocation Volterra elastic field.

Equilibrium properties of a screw dislocation dipole are also
affected by the dislocation core field.
This additional elastic field increases the stability of the dipole:
a higher stress is needed to destroy the dipole. 
Nevertheless, when the two dislocations composing the dipole
are sufficiently separated, one can consider that they only interact
through their Volterra field.

The amplitude of the dilatation corresponding to the dislocation core field
does not depend on the dislocation core configuration,
\ie either easy or hard core structure. 
This indicates that the core field does not arise from the atomic 
structure of the dislocation core, but may be induced by anharmonicity.
Our work, like previous similar studies \cite{HEN04,HEN05,GEH72,HIR73,HOA76,Sinclair1978}, 
shows that such an anharmonic effect can be fully considered 
within linear elasticity theory with the help of a localized core field.
One could have also used non-linear elasticity theory 
to incorporate anharmonic contributions, 
either following the approach of Seeger and Haasen \cite{Seeger1958a}
based on a Grüneisen model for an isotropic crystal, 
or the iterative scheme proposed by Willis \cite{Willis1967,Teodosiu1982}
for an anisotropic crystal.

\begin{acknowledgments}
	This work was supported by EFDA MAT-REMEV programme,
	by the SIMDIM project under contract No. ANR-06-BLAN-250,
	and by the European Commission in the framework of the PERFORM60 project
	under the grant agreement number 232612 in FP7/2007-2011.
	It was performed using HPC resources from GENCI-CINES
	and GENCI-CCRT (Grants No. 2009-096020 and 2010-096020).
\end{acknowledgments}

\appendix

\section{Core energies and core radius}
\label{sec:Ecore_rc}

\begin{figure}[htp] 
  \begin{center}
    \includegraphics[width=0.99\linewidth]{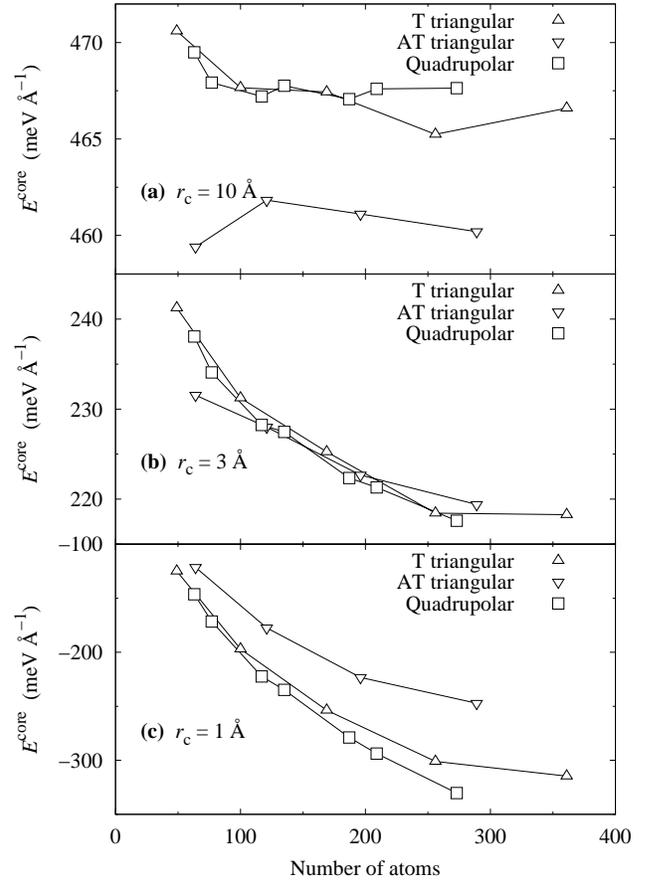}
  \end{center}
  \caption{Core energy of the screw dislocation in the easy core configuration
  for different core radii: 
  (a) $r_{\rm c}=10$\,\AA, (b) $r_{\rm c}=3$\,\AA, and (c) $r_{\rm c}=1$\,\AA.
  Both the Volterra and the core fields have been considered
  in the elastic energy.}
  \label{fig:Ecore_rc}
\end{figure}

The core energy depends on the choice of the core radius $r_{\rm c}$.
This core radius defines the cylindrical region around the dislocation line,
where the strain is so high that elasticity theory does not apply. 
It therefore partitions the dislocation excess energy into two contributions, 
the core energy corresponding to the energy stored in this core cylinder
and the elastic energy in the remaining space.
Changing the value of  $r_{\rm c}$ modifies this partition 
between core and elastic energy without modifying the total excess energy 
[Eq.~(\ref{eq:Edislo})].
In the present work, the choice of $r_{\rm c}$ affects
the convergence of the core energy with the size 
of the simulation box and the geometry of the dislocation periodic array.
This arises from the dislocation core field. 
The dislocation line energy created by this core field depends on $r_{\rm c}$
through the contribution $1/2 \ M_{ij}K_{ijkl}M_{kl} \ 1/{r_{\rm c}}^2$
[Eq.~(\ref{eq:Edislo})]. 
As the dipole moments $M_{ij}$ have been found to depend on the size 
of the simulation cell (Fig.~\ref{fig:dislo_moments}), 
changing the value of $r_{\rm c}$ leads to a shift of the core energy
which also depends on this size.
Figure~\ref{fig:Ecore_rc} shows the core energies obtained for different 
core radii.
One sees that the value $r_{\rm c}=3$\,{\AA} leads to a core energy which does not
depend on the geometry of the dislocation periodic array 
and which converges reasonably with the size of the simulation cell.
Moreover, it is close to the norm of the Burgers vector ($b=2.5$\,\AA)
as theoretically expected\cite{Hirth1982}.

\section{Carbon dipolar tensor}
\label{sec:C}

\begin{figure}[htb]
  \begin{center}
    \includegraphics[width=0.49\linewidth]{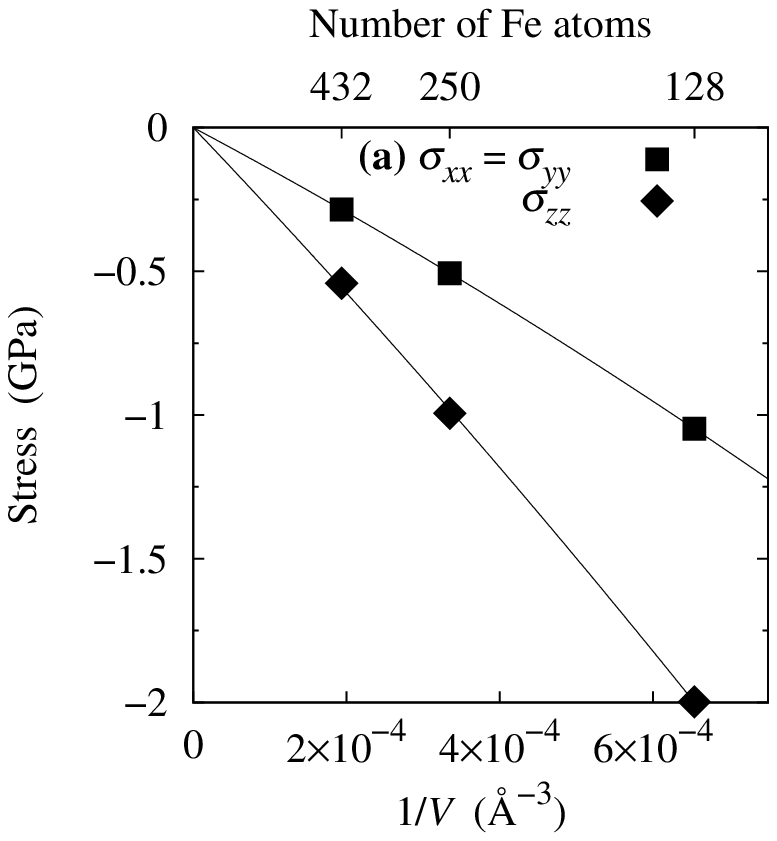}
    \hfill
    \includegraphics[width=0.49\linewidth]{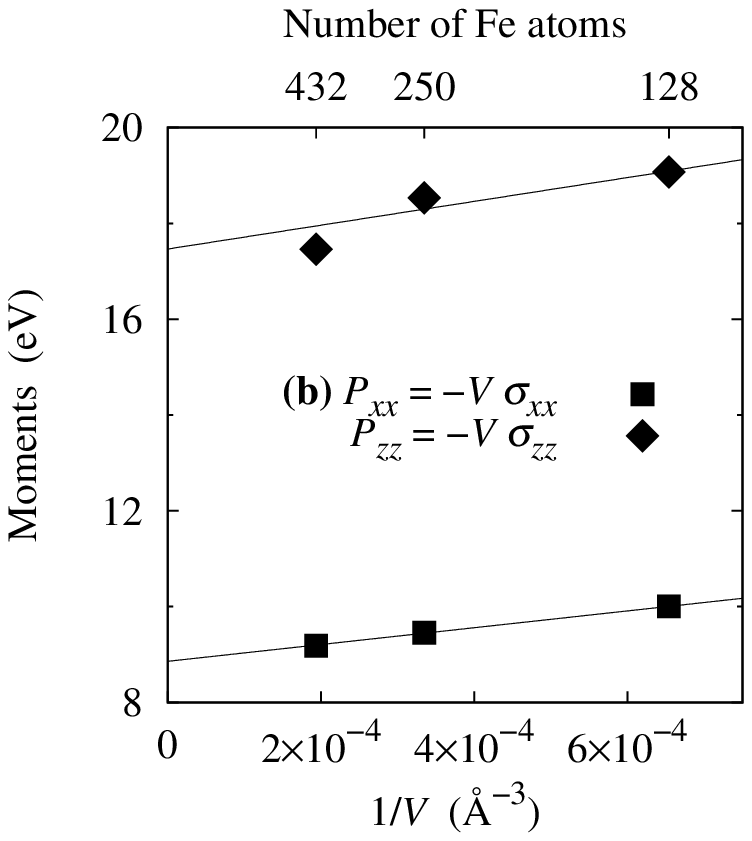}
  \end{center}
  \caption{Variation of the stress $\sigma_{ij}$ 
  and of the corresponding dipolar tensor $P_{ij}$
  with the inverse of the volume $V$ of a unit cell
  containing one C atom in a [001] variant embedded in an Fe matrix.
  Symbols correspond to \emph{ab initio} calculations 
  and solid lines to Eq.~(\ref{eq:C_dipolar_tensor}) 
  and (\ref{eq:C_stress}).}
  \label{fig:C_stress}
\end{figure}

A solute C atom embedded in a Fe matrix is modeled within elasticity theory
by a dipolar tensor\cite{Bacon1980} $P_{ij}$.
As shown in Ref.~\onlinecite{CLO08}, the value of this tensor can be simply 
deduced from the stress tensor measured in atomistic simulations where one solute
atom is embedded in the solvent using periodic boundary conditions. 
One predicts that the homogeneous stress measured in these simulations 
varies linearly with the inverse of the volume $V$ of the unit cell,
\begin{equation*}
  \sigma_{ij} = - \frac{P_{ij}}{V}.
\end{equation*}
Because of the small size of the unit cell used in \emph{ab initio} calculations, 
one has to take into account the polarizability\cite{DED78b,SCH84,PUL86} of the C atom, 
\ie the fact that the tensor $P_{ij}$ depends of the strain $\varepsilon_{kl}^{\rm C}$
locally applied on the C atom. 
One can write to first order
\begin{equation*}
  P_{ij}\left( \varepsilon_{kl}^{\rm C} \right) = P^0_{ij} + P^{1}_{ijkl}\varepsilon_{kl}^{\rm C},
\end{equation*}
where $P^0_{ij}$ is the C dipolar tensor in an unstrained crystal 
and $P^1_{ijkl}$ its first derivative with respect to the applied strain.
In our atomistic simulations,
this applied strain $\varepsilon_{kl}^{\rm C}$ arises from the periodic
images of the C atom.
The strain created by a point defect varies linearly with the inverse of the cube of the separation distance  
and we use for our atomistic simulations homothetic unit cells with the same cubic shape.
The dipolar tensor $P_{ij}$ should therefore vary linearly with the inverse of the volume
of the unit cell,
\begin{equation}
  P_{ij} = P^0_{ij} + \frac{\delta P_{ij}}{V},
  \label{eq:C_dipolar_tensor}
\end{equation}
where $\delta P_{ij}$ is a constant depending only on the shape of the unit cell and not on its volume.
As a consequence, the stress measured in the atomistic simulations should vary like
\begin{equation}
  \sigma_{ij} = -\frac{P^0_{ij}}{V} - \frac{\delta P_{ij}}{V^2}.
  \label{eq:C_stress}
\end{equation}

We performed \emph{ab initio} calculations to obtain the values of 
the dipolar tensor, characterizing one carbon atom in the iron matrix.
We chose a cubic unit cell which contains one C atom 
in an octahedral interstitial site. 
The simulation boxes contain 128, 250, or 432 Fe atoms. 
The \textsc{Siesta} calculation details are the same as for dislocation calculations,
and 13 numerical pseudoatomic orbitals per carbon atom are used to represent
the valence electrons as described in Ref. \onlinecite{Fu2008}.
The k-point grids used for the calculations are $4\times4\times4$ 
for the 128 and 250 atom cells and $3\times3\times3$
for the 432 atom cells.

Because of the tetragonal symmetry of the octahedral interstitial site, 
the dipolar tensor characterizing the C atom is diagonal with only two
independent components. This agrees with the symmetry of the stress tensor
given by \emph{ab initio} calculations. The variations of this stress tensor
with the volume of the unit cell are in perfect agreement with 
Eq.~(\ref{eq:C_stress}) (Fig.~\ref{fig:C_stress}).
This allows us to deduce the elastic dipole $P_{ij}$ which is characterized
by the values $P_{xx}=P_{yy}=8.9$ and $P_{zz}=17.5$\,eV 
for the [100] variant of the C atom in the dilute limit 
[$V\rightarrow \infty$ in Eq.~(\ref{eq:C_dipolar_tensor})].

Previous \emph{ab initio} calculations performed in smaller 
simulation cells have led to the C atom dipolar tensor 
deduced from Kanzaki forces \cite{DOM04}. 
These values are consistent with the ones we have deduced from the homogeneous stress.

\begin{table}[!tbhp]
	\caption{Variations of the iron lattice parameter
	with carbon atomic fraction, $\delta_x$ and $\delta_z$, 
	and formation volume of carbon, $\delta\Omega$.}
	\label{tab:carbon}
	\begin{ruledtabular}
	\begin{tabular}{lccc}
				& $\delta_x$	& $\delta_z$	& $\delta\Omega$ (\AA$^3$) \\
		\hline
		\emph{Ab initio}			& --0.086	& 1.04	& 10.4 \\
		Empirical potential \cite{BEC07} 	& --0.088	& 0.56	& 4.47 \\
		Exp. \cite{COC55}			& --0.052	& 0.76	& 7.63	\\
		Exp. \cite{BAC69}			& --0.0977	& 0.862	& 7.76	\\
		Exp. \cite{CHE90}			& --0.09	& 0.85	& 7.80	\\
	\end{tabular}
	\end{ruledtabular}
\end{table}

The elastic dipole, $P_{ij}$, can be simply related to the parameters $\delta_x$ and $\delta_z$
of the Vegard law \cite{CLO08}, which assumes a linear relation between
the variations of the lattice parameter, $a$ and $c$, and the atomic fraction 
of carbon atoms, $x_{\rm C}$.
If all carbon atoms are located on the $\left[ 001 \right]$ variant of the octahedral site, 
\begin{equation*}
	\begin{split}
	a(x_{\rm C}) &= a_0\left( 1 + \delta_x x_{\rm C} \right) 
		\textrm{, along the $\left[ 100 \right]$ or $\left[ 010 \right]$ axis,} \\
	c(x_{\rm C}) &= a_0\left( 1 + \delta_z x_{\rm C} \right)
		\textrm{, along the $\left[ 001 \right]$ axis,}
	\end{split}
\end{equation*}
where $a_0$ is the pure Fe lattice parameter.
The parameters $\delta_x$ and $\delta_z$ deduced from our DFT calculations
are compared to experimental data \cite{COC55,BAC69,CHE90} in Table~\ref{tab:carbon}.
The \emph{ab initio} calculations lead to a formation volume of carbon
larger than the experimental value, and to a larger tetragonal distortion
expressed as $(\delta_z-\delta_x)$.

One can also deduce from the elastic dipole the variation of the solution enthalpy of C in bcc Fe
with a volume expansion,
\begin{equation*}
  V_0 \left. \frac{\partial H^{\rm exc}}{\partial V} \right|_{V=V_0} =  - \frac{1}{3} P_{ii},
\end{equation*}
where $H^{\rm exc}$ is the excess enthalpy of a Fe crystal of equilibrium volume $V_0$
containing one C atom.
The value corresponding to the elastic dipoles calculated above, $-11.8$\,eV, 
is in good agreement with the value obtained by Hristova \etal \cite{Hristova2011},
$-12.3$\,eV, using an \emph{ab initio} approach based on GGA-DFT with a plane waves basis set 
and the Blöchl projector-augmented wave method (PAW) as implemented 
in the Vienna Ab Initio Simulation Package (VASP). 
This validates our \textsc{Siesta} approach for characterizing C atom 
embedded in an iron bcc matrix.

\bibliographystyle{apsrev4-1}
\bibliography{clouet2011b}

\end{document}